%% file: cav.tex
\documentclass[runningheads]{llncs}
\usepackage{graphicx}
\usepackage{wrapfig,lipsum,booktabs}
\usepackage[english]{babel}
\usepackage[utf8]{inputenc}

\usepackage[fleqn]{amsmath}

\usepackage{mathtools}

\usepackage{amssymb}
\usepackage{stmaryrd}
\usepackage{wasysym}
\usepackage{mathrsfs}
\usepackage{array}

\usepackage{afterpage}

\usepackage[inline]{enumitem}

\usepackage{adjustbox}
\usepackage{caption}

\usepackage{booktabs}
\usepackage{multirow}

\usepackage{soul}

\usepackage{float}

\usepackage{hyperref}
\hypersetup{
    colorlinks=true,
    linkcolor=blue,
    filecolor=magenta,      
    urlcolor=blue,
    citecolor=black
}

\usepackage{framed}

\usepackage{verbatim}

\usepackage{listings}
\usepackage[colorinlistoftodos]{todonotes}

\usepackage{wrapfig}
\usepackage{mdframed}

\input{macros}

\usepackage{algorithm}
\usepackage[noend]{algpseudocode}
\usepackage{microtype}

\def\orcidID#1{\smash{\href{http://orcid.org/#1}{\protect\raisebox{-1.25pt}{\protect\includegraphics{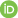}}}}}
\makeatother

\begin{document}

\title{
Deductive Controller Synthesis for Probabilistic Hyperproperties\thanks{This work has been supported by the Czech Science Foundation grant \mbox{GA23-06963S} (VESCAA), the European Union's Horizon 2020 project No 101034440 and by the Vienna Science and Technology Fund 10.47379/ICT19018.
%\begin{wrapfigure}{l}{1cm}\vspace{-5ex}
%\includegraphics[width=1cm]{figures/eulogo.jpg}
%\end{wrapfigure}The project leading to this application has received %funding from the European Union's Horizon 2020 research and innovation %programme under grant agreement No 101034440. This work has been %funded also by the Vienna Science and Technology Fund (WWTF) %[10.47379/ICT19018] 
}
}

\author{Roman Andriushchenko\inst{1}\orcidID{0000-0002-1286-934X} \and Ezio Bartocci\inst{2}\orcidID{0000-0002-8004-6601} \and 
Milan \v{C}e\v{s}ka\inst{1}\orcidID{0000-0002-0300-9727} \and Francesco Pontiggia\inst{2}\orcidID{0000-0003-2569-6238} \and Sarah Sallinger\inst{2}\orcidID{0000-0002-6828-659X}
}

\authorrunning{R. Andriushchenko et al.}
\institute{
Brno University of Technology, Brno, Czech Republic 
\and
TU Wien, Wien, Austria
% \\
%
% \\
% \email{\{iandri,ceskam\}@fit.vutbr.cz, sjunges@berkeley.edu, katoen@cs.rwth-aachen.de}\\
}

%

%
%\institute{Princeton University, Princeton NJ 08544, USA \and
%Springer Heidelberg, Tiergartenstr. 17, 69121 Heidelberg, Germany
\maketitle              % typeset the header of the contribution

\vspace{-1em}
\begin{abstract}
Probabilistic hyperproperties specify quantitative relations between the 
probabilities of reaching different target sets of states from different 
initial sets of states. This class of behavioral properties is suitable for capturing important security, privacy, and system-level requirements. We propose a new approach to solve the controller synthesis problem for Markov decision processes (MDPs) and probabilistic hyperproperties.  Our specification language builds on top of the logic HyperPCTL and enhances it with structural constraints over the synthesized controllers. 
Our approach starts from a family of controllers represented symbolically and defined over the same copy of an MDP. We then introduce an abstraction refinement strategy that can relate multiple computation trees and that we employ to prune the search space deductively.
The experimental evaluation demonstrates that the proposed approach considerably outperforms \textsc{HyperProb}, a state-of-the-art SMT-based model checking tool for HyperPCTL. Moreover, our approach is the first one that is able to effectively combine probabilistic hyperproperties with additional intra-controller constraints (e.g.~partial observability) as well as inter-controller constraints (e.g.~agreements on a common action).
\end{abstract}

% INTRODUCTION
\vspace*{-2em}
\input{sections/1_introduction}

% "BACKGROUND"
\input{sections/2_problem_formulation.tex}

% families & exploration
\input{sections/3-4-exploration.tex}

% EXPERIMENTS
\input{sections/5-eval}

%CONCLUSIONS
\input{sections/conclusions}
\bibliographystyle{splncs04}
\bibliography{biblio.bib}

\appendix
\section{Selecting a conflicting parameter}
\label{ap:metric}

Let $\mathcal{C} \subseteq \mathcal{K}$ denote the set of conflicting parameters in minimizing or maximizing controller $\sigma$.
For the purpose of splitting, we wish to identify a single parameter $k^* \in \mathcal{C}$ that will produce sub-families $\set{\mathcal{R}_X}_{X \in \mathfrak{X}}$ that are more likely to fall into cases \textbf{(1)} or \textbf{(2)}.
Since predicting the exact analysis of resulting sub-families is impossible, we propose the following heuristics based on the results of the model check queries.
For a probability constraint, we introduce the \emph{immediate impact} of action $a$ in state $s$:
\\[-0.5em]
$$\gamma(s,a) \coloneqq exp(s) \cdot \Big[ \sum_{s'} \mathcal{P}(s, \alpha, s') \cdot \reach{M^\sigma,s'}{T} \Big],$$
\\[-0.8em]
where $exp(s)$ is the expected number of visits of state $s$ in MC $M^\sigma$. Likewise, for a reward constraint, we have:
\\[-0.5em]
$$\gamma(s,a) \coloneqq exp(s) \cdot \Big[ rew(s,\alpha) + \sum_{s'} \mathcal{P}(s, \alpha, s') \cdot \expected{M^\sigma,s'}{T} \Big],$$
\\[-0.8em]
where $rew(s,a)$ denotes the reward associated with executing $a$ in $s$. After having computed the immediate impact for each state $s \in S$ and action $a \in Act$,  the parameter with the highest average range of $\gamma$ is selected for splitting:
\\[-0.8em]
$$k^* \in \argmax_{k \in \mathcal{C}}  \avg_{s \in \mathcal{S}_k} \set*{ \max_{a \in \mathcal{A}_k} \gamma(s,a) - \min_{a \in \mathcal{A}_k} \gamma(s,a)}.$$

\section{Short description of the experimental models}
\label{ap:models}

\subsection{Models from \cite{DobeABB21}}
Intuitively, \textbf{TA} and \textbf{PW} both require that for every nondeterministically-chosen value of a secret (also called \emph{high input}), the programs have the same distribution over the set of measurable execution times (the \emph{low output}). This ensures the absence of \emph{timing side channels} that may disclosure information about the sensitive input. On the other hand, \textbf{TS}, expresses the \emph{scheduler-specific probabilistic observational determinism} ($\mathsf{SSPOD}$)~\cite{NgoSH13} hyperproperty: given a probabilistic multi-threaded program, 
and a specific set of probabilistic schedulers under investigation, any pair of executions from two low-equivalent initial states (i.e., states that are indistinguishable with respect to the input secret information) should terminate in low-equivalent states (i.e., produce the same output) with equal probabilities. 
Finally, the \textbf{PC} seeks for a scheduler that induces a DTMC which simulates the outcomes of 6-sided die (specified as a DTMC) using a fair~coin.
The first three case studies are formulated as \emph{verification} problems and thus we consider the dual formulation, i.e.,  \emph{synthesis of a counter-example}.  

We report the results for selected instances of the case studies. Parameter $n$ determines the number of bits representing the secret, and hence the set of observable execution times. In \cite{DobeABB21}, models up to $n=4$ are considered. While replicating the experiments from the original benchmark, we also report on two variants ({PW}$^{*}(n=4)$ and {TA}$^{*}(n=4)$) where the full set of observable outputs is constrained. Despite having the same model, these variants are significantly harder to solve due to a more complex (viz. lengthier) specification. In TS, $h$ is a variable representing the secret. Every instance of TS analyzes the same problem for the given two different values of the secret. In PC, $s$ indicates the states where appropriate actions have to be synthesized. 

\subsection{Models from \textbf{SD} Case Study}
The \texttt{simple} model is the Maze presented as first Case Study in \cite{AndriushchenkoC21}. \texttt{larger-1}, \texttt{larger-2} and \texttt{larger-3} are 5x5 Mazes, equipped with different sets of trap states and different initial states. \texttt{splash-1} and \texttt{splash-2}, resp. 4x4 and 5x5 Mazes, include some states with a high probability of error with respect to the chosen direction. This increases the variance between the success probabilities of different paths, hence it makes the planning problem harder. Finally, in \texttt{train} (7x7) there are some fixed paths that, if taken, forbid the robot to go back.

\end{document}

%% file: macros.tex
\DeclareMathOperator*{\argmax}{arg\,max}

\newcommand{\mpm}{\mathcal{P}}
\newcommand{\mdpT}{(S, Act, \mpm)}

\newcommand{\probability}[1]{\mathbb{P}(#1)}

\newcommand{\reach}[2]{\probability{#1 \models \Diamond #2}}
\newcommand{\expected}[2]{\mathbb{R}({#1 \models \Diamond #2})}

\newcommand{\phistruc}{\varphi^{struc}}
\newcommand{\phistate}{\varphi^{state}}
\newcommand{\phiprob}{\varphi^{prob}}
\newcommand{\phiatom}{\varphi^{atom}}

\newcommand{\sv}[1]{{s_{#1}}}
\newcommand{\cv}[1]{\sigma_{#1}}
\newcommand{\ci}[1]{\sigma^{#1}}

\newcommand{\qsis}[2]{\mathcal{Q}_{#1} \sv{#1} {\in} I_{#1} [#2] }

\newcommand{\schedulerset}{\{\cv{1},\dots,\cv{n}\}}
\newcommand{\schedulertuple}{(\cv{1}, \dots, \cv{n})}
\newcommand{\osigma}{\overline{\sigma}}

\newcommand{\existschedulers}{\exists \, \cv{1}, \ldots, \cv{n}}
\newcommand{\existinitials}{\qsis{1}{\sigma^1} \dots \qsis{m}{\sigma^m}}
\newcommand{\hyperspec}{\existschedulers \in \Sigma^M \colon \phistruc \land \existinitials \colon \phiprob}
\newcommand{\same}[2]{\mathcal{X}_{#1}(#2)}
\newcommand{\obs}[2]{\mathcal{O}_{#1}(#2)}

\newcommand{\kp}[2]{k(#1,#2)}
\newcommand{\dom}[2]{\mathrm{dom}(#1,#2)}
\DeclarePairedDelimiter{\set}{\{}{\}}
\DeclareMathOperator*{\avg}{avg}
\newcommand{\oneton}{\{1,\ldots,n\}}
\newcommand{\lb}[1]{l_{#1}}
\newcommand{\ub}[1]{u_{#1}}

%% file: sections/1_introduction.tex
\section{Introduction}
Controller synthesis in a probabilistic environment~\cite{baier2004} is a well-known and highly addressed problem. 
There exist several off-the-shelf tools (e.g.~PRISM~\cite{KwiatkowskaNP11} or STORM~\cite{HenselJKQV22}) that are able to to synthesize optimal policies for large Markov decision processes (MDPs) with respect to probability measures over a given set of paths.
However, recent studies~\cite{AbrahamB18,AbrahamBBD20,DOBE2022104978,DobeWABB22,DimitrovaFT20} have brought up the necessity to investigate a new class of properties that relate probability measures over different sets of executions.
This class of requirements is called \textbf{probabilistic hyperproperties}~\cite{AbrahamBBD20}, and standard logics, such as Probabilistic Computation Tree Logic (PCTL)~\cite{HanssonJ94}, are not expressive enough for them. Classical examples include security and privacy requirements, such as quantitative information-flow~\cite{KopfB07}, probabilistic noninterference~\cite{GrayS98,ONeillCC06} and differential privacy for databases~\cite{Dwork06}.
Probabilistic hyperproperties can be expressed using  Probabilistic HyperLogic (PHL)~\cite{DimitrovaFT20} and HyperPCTL~\cite{AbrahamBBD20}. Both are probabilistic temporal logics that extends PCTL with i) quantifiers over schedulers and ii) predicates over functions of probability expressions. 
Verification of HyperPCTL or PHL formulae can be formalized as the synthesis of multiple, correlated strategies that represent a witness or a counterexample to the specification.
Furthermore, pioneering studies have shown that probabilistic hyperproperties have novel applications in robotics and planning \cite{DimitrovaFT20}, and in causality analysis for probabilistic systems \cite{BaierFPZ22}. %, where the (high) probability of an event is linked with a cause for it. 
To deal with these new constraints, it is necessary to establish a connection not only between the probabilities of different strategies, but also between the choices they make in corresponding states, i.e. reasoning about the \textbf{structural characteristics} of the resulting controllers. Finally, the study of probabilistic hyperproperties in a \textbf{partially observable} setting is mostly unexplored.%, where existing works (such as \cite{RossIMB09,MareckiSV10,Novakovic019}) have only targeted specific Case Studies and not introduced an overall formalization.

\vspace{-0.5em}
\paragraph{Motivating Example 1.} \label{ex:motivating_example}
\begin{wrapfigure}[8]{r}{0.35\textwidth}
    \centering
    \vspace{-2.5em}
    \includegraphics[scale = 0.20]{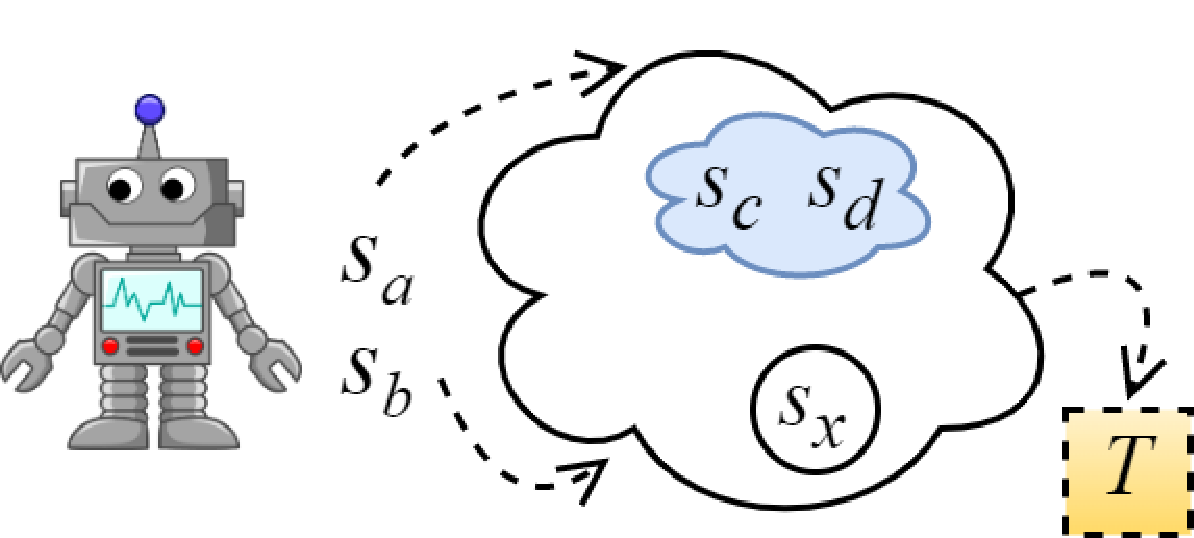}
    \caption{Planning under probabilistic noninterference.}
    \label{fig:robotic_example}
\end{wrapfigure}
Consider a planning problem depicted in Figure~\ref{fig:robotic_example}: a robot in a probabilistic environment attempts to reach the target location~$T$.
A controller for the robot is a program that, for each position, 
%(or observation in the context of POMDPs)
selects the next action to take.
Consider the synthesis of two distinct controllers for two possible initial states $s_a$ and $s_b$.
In a confidential setting, we want to combine performance constraints on success probabilities
% (where good but not necessarily optimal policies are requested)
with additional privacy and structural requirements.
\emph{Probabilistic noninterference} is a hyperproperty asserting that the success probability must be the same for $s_a$ and $s_b$, so that an attacker cannot infer any information about the selected initial location by observing the percentage of successful robots.
If further the choice made in some state $s_{\chi}$ is assumed to be observable to the attacker, the two strategies must agree on their action choices in $s_{\chi}$. Finally, the \emph{partial observability} constraint requires a controller to make the same choice in states $s_c$ and $s_d$ having the same~observation.

\vspace{-.5em}
\paragraph{Related Work.} In~\cite{AbrahamB18} and~\cite{AbrahamBBD20_LPAR}, the authors introduce, respectively, the model-checking problem and the parameter synthesis problem for discrete-time Markov chains satisfying probabilistic hyperproperties.
The works in~\cite{AbrahamBBD20,DimitrovaFT20} have recently studied the verification of probabilistic hyperproperties in the context of MDPs and proposed two suitable logics: HyperPCTL and Probabilistic HyperLogic (PHL), respectively. However, in both cases the model-checking problem is undecidable. 
The solution proposed for the former logic is to consider only the class of deterministic memoryless controllers, where the problem is shown to be NP-complete by a suitable SMT encoding. Despite recent improvements of \textsc{HyperProb}~\cite{DobeABB21,DOBE2022104978}, a state-of-the-art tool implementing this approach, its scalability remains very limited: the complexity of the encoding gets out of reach even for modest models. For instance, a simple side-channel analysis introduced in~\cite{AbrahamBBD20} (checking that the probability of certain outcomes is equal) on an MDP with 244 states leads to an SMT problem with over a million variables, which \textsc{HyperProb} fails to solve.
%$1204628$ variables and  $489$ constraints. 
% these numbers come from TA.4* (n=4) experiment %%
Finally, \textsc{HyperProb} does not provide support for structural constraints describing non-probabilistic relations between controllers, and the presented techniques cannot principally handle neither partial observability. %, despite the work of~\cite{DOBE2022104978} mentions to have solved one model with partial observability.

On the other hand, PHL is able to express such structural constraints using HyperCTL*~\cite{ClarksonFKMRS14} formulae in conjunction with probability expressions. Hence, with PHL, it is possible to require different paths to take the same (or a different) choice in a certain state. However, the model checking procedure developed in~\cite{DimitrovaFT20}  employs a bounded synthesis algorithm which supports only \emph{qualitative} constraints, and cannot effectively handle the probability expressions of a PHL formula - the synthesis loop has to check them a posteriori. As a result, when the qualitative constraints do not restrict the controllers sufficiently,  a large part of the search space has to be explicitly enumerated. 

Various works have addressed (reactive) synthesis problems in non probabilistic settings \cite{BonakdarpourF20,0044NP20,FinkbeinerHLST18,FinkbeinerHLST20}. In particular,  decidability for HyperLTL~\cite{ClarksonFKMRS14} specifications and finite-state plants has been shown in~\cite{BonakdarpourF20}.

\vspace{-0.3em}
\paragraph{Our contribution.} To mitigate the limitations of the existing approaches, we propose a novel controller synthesis framework for MDPs and probabilistic hyperproperties. We enrich expressive power of HyperPCTL with \textbf{structural constraints} to support additional requirements on the controllers (Section~\ref{sec:background}).
We treat the synthesis problem as an exploration problem over a \textbf{symbolic representation} of a family of controllers defined over the original MDP (Section~\ref{sec:family}). 
%Such approach naturally extends to the \emph{optimal synthesis}, or a more complicated and unexplored class of problems such POMDP controller synthesis and the synthesis of finite-state controllers. Note that, due to the general undecidability of the problem, this matter is relevant even in the context of simple MDPs. 
Inspired by recent advances in the synthesis of probabilistic systems~\cite{Andriushchenko021}, we tailor a deductive exploration method to support hyper-reasoning. In particular, we propose an \textbf{abstraction refinement strategy} that relates multiple computation trees and allows us to deductively identify sets of violating controllers that can be safely removed from the search space (Section~\ref{sec:exploration}). A detailed experimental evaluation (Section~\ref{sec:eval}) demonstrates that our approach considerably outperforms the state-of-the-art tool \textsc{HyperProb}~\cite{DobeABB21}. In particular, our approach scales to more complicated synthesis problems where \textsc{HyperProb} times out, and supports, for the first time, various structural constraints.

%% file: sections/2_problem_formulation.tex
\vspace{-0.4em}
\section{Problem Formulation} \label{sec:background}
In this section we present the formalisms involved in our synthesis framework as well as the problem statement. We refer to \cite{BaierAFK18,0020348} for further background.
\vspace{-0.5em}
\subsection{Probabilistic models}
A (discrete) \emph{probabilistic distribution} over a finite set $X$ is $\mu \, : X \rightarrow [0,1]$, where $\sum_{x \in X} \mu(x) = 1$. The set of all distributions on $X$ is denoted as $Distr(X)$. 

% MDP
\begin{definition}[MDP]
A \emph{Markov Decision Process (MDP)} is a tuple $\mdpT$ with a finite set $S$ of states, a finite set $Act$ of actions and a transition function $\mpm: S \times Act \rightarrow Distr(S)$. An MDP with $|Act| = 1$ is a \emph{Markov chain (MC)} $(S,\mpm)$ with $\mpm: S \rightarrow Distr(S)$. 
\end{definition}
We depart from the standard notion of MDP~\cite{0020348} and do not specify initial states: those will be derived from the context given by the specification.
Additionally, to simplify the presentation, we enable all actions in every state.
An MDP can be equipped with a state-action reward (or cost) function $rew: S {\times} Act \rightarrow \mathbb{R}_{\geqslant 0}$, where reward $rew(s,\alpha)$ is earned upon executing action $\alpha$ in state~$s$.
A (memoryless, deterministic) \emph{controller} $\sigma \colon S \rightarrow Act$ selects in each state $s$ an action $\sigma(s)$; let $\Sigma^M$ denote the set of all such controllers in $M$.
Imposing controller $\sigma$ onto MDP $M = \mdpT$ yields MC $M^\sigma = \left(S,\mpm^\sigma\right)$ where \mbox{$\mpm^\sigma(s) = \mpm(s,\sigma(s))$}.
For an MC $M$, $\reach{M,s}{T}$ denotes the probability of reaching the set $T \subseteq S$ of target states when 
starting in state $s \in S$. If \mbox{$\reach{M,s}{T} = 1$}, then $\expected{M,s}{T}$ denotes the expected reward 
accumulated from $s$ before reaching~$T$. Using the standard techniques, one can effectively compute a minimizing 
controller $\sigma_{\min}$ that minimizes the probability to reach~$T$, i.e. $\forall \sigma \in \Sigma^M:$ 
\mbox{$\reach{M^{\sigma_{\min}},s}{T}$} $\leq \reach{M^{\sigma},s}{T}$.
Naturally, this can be lifted to the expected rewards.
% Let $\mathbb{P}_{\min}(M, s \models 
% \Diamond T)$ denote the minimal probability in the MDP $M$. Similarly, we can compute a maximizing controller 
% $\sigma_{\max}$ with the maximal probability  $\mathbb{P}_{\max}(M, s \models \Diamond T)$.

% The \emph{available actions} in $s \in S$ are $Act(s) = \{ a \in Act \mid \lltrans(s,a) \neq \bot \}$.

%A \emph{path} of an MDP $M$ is an (in)finite sequence $\pi = s_0 \xrightarrow[]{a_0} s_1 \xrightarrow[]{a_1} \dots$ 
%where $s_0 \in I$, $s_i \in S$, $a_i \in Act(s_i)$, and $\mathcal{P}(s_i, a_i)(s_{i+1}) \neq 0$ for all $i \in \mathbb{N}$. 
%For finite $\pi$, $last(\pi)$ denotes the last state of $\pi$. The set of (in)finite paths of $M$ is $Paths^{M}_{fin}$($Paths^{M}_{}$)

% A \textup{Markov Chain (MC) induced by MDP $M := (S, I, Act, \lltrans)$ and controller $\sigma \in \Sigma_{M}$} is given by $M^{\sigma} := (S, s_0,\lltrans^{\sigma})$ where $\lltrans^{\sigma}(s,s') := \lltrans(s, \sigma(s))(s')$

\subsection{HyperPCTL with Structural Constraints}
\label{sec:spec}

Assume an MDP $M = \mdpT$. Our specification language builds on the syntax of HyperPCTL~\cite{AbrahamBBD20}, enriched by structural constraints. For both principal and technical reasons discussed below, we support only the following fragment of the original hyperlogic:

\vspace{-1em}
\begin{alignat*}{3}
& \varphi  && \coloneqq \existschedulers \in \Sigma^M \colon \phistruc \land \phistate  \\
&\phistruc && \coloneqq \phistruc \land \phistruc \mid \same{s}{\Sigma} \mid \obs{S'}{\sigma_i} \\
%\diff{\bowtie k}{\cv{i}}{\cv{j}} \\
%
&\phistate   && \coloneqq \existinitials \colon \phiprob \\
&\phiprob    && \coloneqq \neg \phiprob  \mid \phiprob \land \phiprob \mid \phiatom \\
&\phiatom    && \coloneqq
\reach{M^{\ci{i}},\sv{i}}{T} \bowtie \lambda \mid  \reach{M^{\ci{i}},\sv{i}}{T_i} \bowtie  \reach{M^{\ci{j}},\sv{j}}{T_j} \mid \\
& & & \hspace{1.7em} \expected{M^{\ci{i}},\sv{i}}{T} \bowtie \rho \mid \expected{M^{\ci{i}},\sv{i}}{T_i} \bowtie \expected{M^{\ci{j}},\sv{j}}{T_j}
\end{alignat*}
where $\mathcal{Q} \in \{ \exists, \forall \}$, $\bowtie \, \in {\{<,\leq,=,>,\geq\}}$, $k \in \mathbb{N}$, $\lambda \in [0,1]$ and $\rho \in \mathbb{R}_{\geq 0}$. The individual parts of the syntax and semantics are described below.

\begin{description}[leftmargin=0pt,itemsep=0.5em]
\item[Controller quantification ($\varphi$):] We consider the \emph{synthesis} problem, i.e. only the existential quantification over the controllers $\cv{i}$ is supported.\footnote{A \emph{verification} problem defined using universal quantification can be treated as synthesis of a counterexample. Alternation in the controller quantification is not supported as it principally complicates the inductive synthesis strategy. To the best of our knowledge, only very few techniques so far practically support alternation~\cite{HsuBKT22,BeutnerF22}. %While PHL and HyperPCTL allow it, they do not manage to support it in the practice, and the presented model checking techniques are restricted to single-controller quantification (for HyperPCTL) and to alternation-free fragments of the logic (for PHL).
} 
    
\item[Structural constraints ($\phistruc$):] In contrast to HyperPCTL~\cite{AbrahamBBD20}, we support conjunctions of structural constraints of two types:
i) intra-controller constraint $\obs{S'}{\sigma_i}$ requiring that the controller $\sigma_i$ takes the same action in all states from the set $S' \subseteq S$ (this can effectively encode partial observability) and 
ii) inter-controller constraint $\same{s}{\Sigma}$ requiring that all controllers in $\Sigma \subseteq \schedulerset$ take the same action in state $s \in S$.

%ii) constraint $\diff{\bowtie k}{\cv{i}}{\cv{j}}$ requiring that $|\{s {\in} S \mid \cv{i}(s) \neq \cv{j}(s)\}| \bowtie k$, i.e.~the number of different choices taken by variables $\cv{i}$ and $\cv{j}$ in individual states meets $\bowtie k$.

\item[State quantification ($\phistate$):] Allows to quantify over the computational trees rooted in the given initial states. We support an arbitrary quantification $\qsis{i}{\ci{i}}$, where each $\sv{i}$  is associated with its domain $I_i \subseteq S$ and some controller $\ci{i} \in \schedulerset$. Note that multiple state variables can be associated  with the same controller ($n \leq m$). 
    
\item[Probability constraints ($\phiprob$):] 
We consider a Boolean combination
of the standard inequalities between the reachability probability (reward) and a threshold  as well as the inequalities that correspond to the class of 2-hyperproperties~\cite{ClarksonS10}, where a relationship between two computation trees is considered\footnote{An extension to nested-free PCTL path formulae is straightforward. Principally, we can extend the probability constraints to allow more complicated expressions (e.g. sum/difference of reachability probabilities from different states) eventually requiring reasoning about more than two computation trees.}.

%$\reach{\ci{i},\sv{i}}{T}$, or reward expressions of type $\expected{\ci{i},\sv{i}}{T}$, where $\sv{i}$ is an in-scope state variable and $T \subseteq S$. In addition to standard PCTL inequalities of the form $\reach{\ci{i},\sv{i}}{T} \bowtie \lambda$, they can also be of type $\reach{\cv{i},\sv{i}}{T_i} \bowtie  \reach{\cv{j},\sv{j}}{T_j}$ relating two reachability (or reward) expressions with different state variables $\sv{i}$ and $\sv{j}$. These inequalities correspond to the class of 2-hyperproperties~\cite{ClarksonS10}, where relationships between two computation trees are considered\footnote{An extension to nested-free PCTL path formulae is straightforward. Principally, we can extend the probability constraints to allow a more complicated expressions (e.g. sum/difference of reachability probabilities from different states) eventually requiring reasoning about more than two computation trees.}.
\end{description}

\vspace{-0.5em}
To evaluate whether given $n$ controllers for MDP $M$ 
satisfy a state formula $ \phistate $, we instantiate the initial states on the probabilistic constraint $\phiprob$. 

\vspace{-0.2em}
\begin{definition}[Instantiation of state quantifiers]
\label{def:inst}
Assume a state formula $\existinitials \colon \phiprob$. Each quantifier $\qsis{i}{\sigma^i}$ is instantiated on $\phiprob$ by substituting $\phiprob$
with ${\nabla_{s \in I_i}\phiprob[\sv{i} \mapsto s]}$, where $\nabla \equiv \bigvee$ if $\mathcal{Q}_i \equiv \exists$ and $\nabla \equiv \bigwedge$ otherwise. 
\end{definition}
\vspace{-0.2em}
Upon instantiation, we obtain a Boolean combination of  inequalities over probabilities and rewards in the corresponding MCs having the given initial states.

% \paragraph{Satisfiability of the state formula}
% To evaluate whether $n$ controllers $\sigma_1,\ldots \sigma_n$ of an MDP $M$ 
% satisfy a state formula 
% $ \phistate \equiv \existinitials \colon \phiprob $ 
% we need to instantiate $\phiprob$. For each state quantifier $\qsis{i}{\ci{i}}$ we instantiate $\phiprob$ with 
% with ${\nabla_{s \in I_i}\phiprob[\sv{i} \mapsto s]}$, where $\nabla \equiv \bigvee$ if $\mathcal{Q}_i \equiv \exists$ and $\nabla \equiv \bigwedge$ otherwise. Upon instantiation, we obtain a Boolean combination of  inequalities over probabilities and rewards in corresponding MC $M^{\ci{i}}$ with initial states $s\in I_i$.

\vspace{-0.5em}
\paragraph{Remark 1.} HyperPCTL syntax also allows to reason about stepwise \emph{paired} multiple traces by associating the atomic propositions in the path formulae of probability expressions with state variables (i.e the initial state of a particular computation). Then, to evaluate a formula with $m$ state quantifications, the semantics requires to build the $m$-cross-product (also called \emph{self-composition} \cite{AbrahamBBD20,DimitrovaFT20}, or \emph{trivial coupling}~\cite{barthe_hsu_2020}), of the original model $M$. In this paper, for the sake of simplicity, we avoid such an involved construction and consider only single, uncoupled computation trees, where all atomic propositions share the same state variable specified in the probability expression. However, handling such constraints would require only an additional technical step performing the cross-product when needed \cite{ZamanCAB22}, and no principle obstacles.

\paragraph{Problem statement.}
The key problem statement in this paper is \emph{feasibility}:
\begin{mdframed}[backgroundcolor=blue!10,innerleftmargin=0.5em,innerrightmargin=0.5em]
Given an MDP $M$ and a hyperspecification  
\vspace{-0.5em}
$$\hyperspec$$
find controllers $\sigma_1,\ldots,\sigma_n$ in $\Sigma^M$ that satisfy $\phistruc$ and the instantiated $\phiprob$.
\end{mdframed}

\paragraph{Remark 2.}
We follow existing approaches~\cite{AbrahamBBD20,DobeWABB22,DOBE2022104978} and consider only the class of deterministic memoryless controllers where synthesis is decidable. These controllers are, however, too weak for some synthesis problems considered in the experimental evaluation. Therefore, we enhance the underlying MDPs with a memory to explore a richer class of controllers, as in~\cite{DimitrovaFT20}.

\paragraph{Example 2.}
Consider Example~\ref{ex:motivating_example} from the introduction. The probabilistic noninterference specification for the synthesis of two controllers can be expressed as
\vspace{-1em}
\begin{alignat*}{1}
\exists \, \cv{1}, \cv{2} &\colon \phistruc \land \forall \sv{1} \in \{s_a\}[\cv{1}] \; \forall \sv{2} \in \{s_b\}[\cv{2}] \colon \phiprob \mbox{, where } \\ 
\phistruc &\equiv \same{s_\chi}{\{\cv{1},\cv{2}\}} \land \obs{\set{s_c,s_d}}{\sigma_1} \land \obs{\set{s_c,s_d}}{\sigma_2} \\
\phiprob &\equiv \reach{M^{\cv{1}},\sv{1}}{T} = \reach{M^{\cv{2}},\sv{2}}{T} \land \reach{M^{\cv{1}},\sv{1}}{T} \geq \lambda 
\end{alignat*}
Given $\cv{1}$ and $\cv{2}$, the instantiation of the state quantifiers on $\phiprob$ yields the following probability constraint $\reach{M^{\cv{1}},s_a}{T} = \reach{M^{\cv{2}},s_b}{T} \land \reach{M^{\cv{1}},s_a}{T} \geq \lambda$. 

\paragraph{Remark 3.}
Our approach also supports various variants of \textbf{optimal synthesis} problems, where an additional minimizing/maximizing objective over a reachability probability or reward is considered. For example, we can maximize the difference $\reach{M^{\cv{1}},\sv{1}}{T_1} - \reach{M^{\cv{2}},\sv{2}}{T_2}$.
%or minimize the distance $\lldiff(\sigma_1,\sigma_2)$ 
%between two controllers, i.e.. 
To simplify the technical exposition of the paper, we will omit the precise formalisation of these problems.

%% file: sections/3-4-exploration.tex
\section{Family of Controllers}
\label{sec:family}

% Since $m$ state quantifications consider a sequence $(\induced{i}{j})_{i=1}^m$ of $m$ Markov chains, where each $\sigma_{i}$ is a controller of $\osigma$, we wish to design a technique that will allow us to efficiently explore this family of $m$-tuples of MCs and identify the chain that encodes solution to the hyperspecification . To reason about $m$ computation trees of $M$ at once, we introduce the \emph{$m$-replication}: an asynchronous composition of multiple copies of $M$.

% \begin{definition}[$m$-replication]
% Let $M = \mdpT$ be an MDP and let $M_{i} = (S_i, Act, \mpm_i)$ denote its \emph{$i$-th replica} with the renamed set $S_i \coloneqq \{s_i \mid s \in S\}$ of states and transition  function $\mpm_i(s_i, \alpha)(s_i') \coloneqq \mpm(s,\alpha)(s')$.
% An \emph{$m$-replication} of $M$ is the MDP $\rmdp = (\mathcal{S}, Act, \rmpm) \coloneqq \bigcup_{i=1}^m M_i$, i.e.~$\mathcal{S} \coloneqq \bigcup_{i=1}^m S_i$ and $\rmpm(s_i,\alpha) \coloneqq \mpm_i(s_i,\alpha)$.
% States $s_i$ and $s_j$ of $\rmdp$ are called \emph{matching}, if they represent the same state $s$ of $M$ in replicas $M_i$ and $M_j$.
% \end{definition}

Consider an MDP $M$ for which we wish to synthesize $n$ controllers $\sigma_1,\ldots \sigma_n$, further referred to as an \emph{$n$-controller}
$\osigma = \schedulertuple$, that satisfy a hyperproperty $\phistruc \land \phistate$.
Motivated by the inductive synthesis method presented in~\cite{Andriushchenko021}, we base our approach on a symbolic representation of a \emph{family of $n$-controllers} over MDP $M$.
The following definition allows us to efficiently represent all $n$-controllers that satisfy $\phistruc$.

% Our starting point will be the following definition of a family of $i$-th controllers that satisfy \emph{intra-controller constraints} of type $\obs{S'}{\sigma_i}$.

\begin{definition}[Parameter space]
\label{def:parameters}
Assume an MDP $M = \mdpT$ and a~hyperspecification $\existschedulers \colon \phistruc \land \phistate $. For each $i \in \oneton$ and $s\in S$, we introduce a parameter $\kp{i}{s}$ with domain $\dom{i}{s} = Act$. Relation $k \simeq k'$ denotes that parameters $k,k'$ are \emph{synonymous}, i.e.~$k$ and $k'$ are different names for the same parameter.
For each  $\obs{S'}{\sigma_i} \in \phistruc $ we require that  \mbox{$\forall \, s,s'\in S':$} $\kp{i}{s} \simeq \kp{i}{s'}$.
Additionally, for each $\same{s}{\Sigma} \in \phistruc $ we require that \mbox{$\forall \, \sigma_i, \sigma_j \in \Sigma :$}  $\kp{i}{s} \simeq \kp{j}{s}$.
Let $\mathcal{K}$ denote the set of all parameters.
\end{definition}
Parameter $\kp{i}{s}$ essentially encodes possible action selections available to~$\sigma_i$ at state $s$. A realisation then encodes a specific assignment of these parameters.

\begin{definition}[Realisation]
\label{def:realisation}
A \emph{realisation} is a mapping $r \colon \mathcal{K} \rightarrow Act$ 
%\fra{This such that... may be considered superfluous because we already set the domain to Act in the previous definition. It makes sense only in the light of subsequent splits, but we haven't introduced them yet.}
%\milan{This okay.}
such that $r(\kp{i}{s}) \in \dom{i}{s}$.
% $\ceil{r}$ denotes arbitrary total realisation that completes $r$, i.e.$r(\kp{i}{s}) \in \dom{i}{s} \Rightarrow \ceil{r}(\kp{i}{s}) = r(\kp{i}{s})$.
$\mathcal{R}$~denotes the set of all realisations.
Realisation $r$ \emph{induces} $n$-controller $\osigma^r \coloneqq (\sigma_1^r,\ldots,\sigma_n^r)$ where $\forall i \in \oneton \, \forall s \in S \colon \sigma_i^{r}(s) = r(\kp{i}{s})$.
\end{definition}
% \emph{The design space} is the set $\mathcal{D} \coloneqq \prod\{\dom{i}{s} \mid i \in \oneton, s \in S\}$ of parameter domains.

Realisation $r \in \mathcal{R}$ specifies for each $\sigma_i$ a specific action $r(\kp{i}{s})$ to be selected in state $s$, and thus describes an $n$-controller $\osigma^r \in (\Sigma^M)^n$.
Using synonymous parameters implies that $r$ always induces controller $\osigma^r$ satisfying structural constraints $\phistruc$.
Thus, the set $\mathcal{R}$ of realisations directly corresponds to the family $\mathcal{F}$ of controllers that satisfy $\phistruc$.
%This construction is formalised in the following definition.

\begin{definition}[Family of controllers]
\label{def:family}
A \emph{family of $i$-th controllers} is the set $\mathcal{F}_i = \set{ \sigma_i^{r} \mid r \in \mathcal{R}  }.$
A \emph{family of $n$-controllers} is the set $\mathcal{F} \coloneqq \set{ \osigma^{r} \mid r \in \mathcal{R}  }$.
%= \prod_{i \in \oneton} \mathcal{F}_i$.
\end{definition}

From this definition it is clear that $\mathcal{F} \subseteq (\Sigma^M)^n$ and that $\osigma \in \mathcal{F} \Leftrightarrow \osigma \models \phistruc$.
We say that an $n$-controller $\osigma = \schedulertuple$ from the set $(\Sigma^M)^n {\setminus} \mathcal{F}$ is \emph{inconsistent}, where we distinguish between the following two types of inconsistency.

\vspace{-0.5em}
\begin{itemize}
\item $\osigma \not \models \obs{S'}{\sigma_i}$ is said to be \emph{intra-controller inconsistent}: controller $\sigma_i$ diverges in its choices $\sigma_i(s_a) \neq \sigma_i(s_b)$ for the action selection in states $s_a,s_b \in S'$.
\item $\osigma \not \models \same{s}{\Sigma}$ is said to be \emph{inter-controller inconsistent}: controllers $\sigma_i,\sigma_j \in \Sigma$ diverge in their choices $\sigma_i(s) \neq \sigma_j(s)$ for the action selection in state $s$.
\end{itemize}
Conversely, if $\osigma \in \mathcal{F}$, then there exists exactly one realisation $r \in \mathcal{R}$~s.t. $\osigma^r = \osigma$. Subsequently, if $\sigma_i \in \mathcal{F}_i$, then the following set of realisations:
\\[-0.5em]
$$\mathcal{R}[\sigma_i] \coloneqq \set*{r \in R \mid \osigma^r = (\sigma_1,\ldots,\sigma_i,\ldots,\sigma_n)}$$
is not empty and contains all realisations that induce $\sigma_i$.
From $\mathcal{F} \subseteq (\Sigma^M)^n$ it follows that the size of the family $|\mathcal{F}| \leq |Act|^{|S|n}$ is exponential in the number~$|S|$ of states and~$n$, rendering enumeration of each $n$-controller $\osigma\in \mathcal{F}$ infeasible.

\section{Deductive Exploration}
\label{sec:exploration}
We present a deductive method allowing for an efficient exploration of the given family $\mathcal{F}$ of $n$-controllers.
We adapt the abstraction refinement approach for PCTL specifications presented in~\cite{CeskaJJK19,Andriushchenko022} to the synthesis against probabilistic hyperproperties.
We explain the method on a canonical synthesis problem $\varphi_c \equiv \exists \, \sigma_1, \sigma_2: \phistruc \land \exists \, s_1 \! \in \! \{s_1\}[\sigma_1], \, \exists s_2 \!\in\! \{s_2\}[\sigma_2]: \phiprob$, where $\phiprob \equiv \reach{M^{\cv{1}},\sv{1}}{T_1} \leq \reach{M^{\cv{2}},\sv{2}}{T_2}$.
% It thus requires to find a 2-controller $\osigma = (\sigma_1,\sigma_2)$ satisfying a single probability constrain $\phiprob$.
A generalisation to more complicated specifications is discussed in Subsection~\ref{subsec:algorithm}. 

% AR
\subsection{Guided Abstraction Refinement Scheme}
\label{subsec:ar}

MDP $M$ and the corresponding set $(\Sigma^M)^n$ of controllers provide an over-approxi\-mation of the set $\mathcal{F}$ of admissible controllers.
The key idea is to analyse~$M$ using standard techniques to obtain the following four controllers: $\sigma_1^{\min}$ ($\sigma_1^{\max}$) minimising (maximising) the probability of reaching $T_1$ from $s_1$, and controllers $\sigma_2^{\min}$ and $\sigma_2^{max}$ minimising (maximising) the probability of reaching~$T_2$ from~$s_2$.
Let $\lb{1} \coloneqq \reach{M^{\sigma_1^{\min}},s_1}{T_1}$ denote the minimum probability of reaching $T_1$ from $s_1$ via $\sigma_1^{\min}$; since the corresponding $\sigma_1^{\min}$ is not necessarily in~$\mathcal{F}_1$, then this probability is a lower bound on $\min_{\sigma \in \mathcal{F}_1} \reach{M^{\cv{}},s_1}{T_1}$. Let $\ub{1}$ be the corresponding upper bound on the probability of reaching $T_1$ from $s_1$, and similarly for $\lb{2}$ and $\ub{2}$. It is clear that $\lb{1} \leq \ub{1}$ and $\lb{2} \leq \ub{2}$.

Comparing the corresponding numerical intervals $[\lb{1},\ub{1}]$ and $[\lb{2},\ub{2}]$ may allow us to %draw specific conclusions 
determine whether $(\Sigma^M)^n$ (and, by extension, $\mathcal{F}$) can satisfy $\phiprob$. The procedure is summarised in Table~\ref{tab:ar_reasoning}. %, where all cases are considered. 
For instance, if the two intervals do not overlap, then %it must hold that 
either all controllers satisfy $\phiprob$ (case \textbf{(1)}) or none do (case \textbf{(2)}).

In the remaining cases \textbf{(3)}-\textbf{(5)}, when the intervals do overlap, it is ambiguous whether $\mathcal{F}$ contains feasible solutions, so additional analysis is necessary.
For example, if $\lb{1} \leq \lb{2}$ (case \textbf{(3)}) and $\sigma_1^{\min}$ satisfies all intra-controller constrafints $\obs{S'}{\sigma_1} \in \phistruc$, then $\sigma_1^{\min} \in \mathcal{F}_1$ and thus there exists realisation $r \in \mathcal{R}[\sigma_1^{\min}]$ such that $\osigma^r \models \varphi_c$.
Analogous reasoning can be applied to case \textbf{(4)}. Finally, in case \textbf{(5)}, an $n$-controller $(\sigma_1^{\min},\sigma_2^{\max})$ represents a solution if it is consistent; then, any realisation $r \in \mathcal{R}[\sigma_1^{\min}] \cap \mathcal{R}[\sigma_2^{\max}]$ that induces both of these controllers is accepting.
In all of these cases \textbf{(3)}-\textbf{(5)}, if the consistency requirements are not met, $\mathcal{R}$ may still contain feasible solutions, but not involving $\sigma_1^{\min}$ or $\sigma_2^{\max}$.
In other terms, our over-approximation $M$ is too coarse and needs to be \emph{refined} to obtain more precise results.
% We do this by partitioning the set $\mathcal{R}$ of realisations into smaller subfamilies disabling actions leading to inconsistencies. % okay the rephrasing
Refinement will be the focus of Subsection~\ref{subsec:splitting}.

\begin{table}[t]
    \centering
    \begin{tabular}{ c | c | c | c }
    \hline
    \textbf{case} & condition & feasibility & SAT realisations \\\hline
    \textbf{(1)} & $[\lb{1}, \ub{1}]  \leq [\lb{2},\ub{2}]  $ & all SAT  & $\mathcal{R}$\\
    \textbf{(2)} & $[\lb{2}, \ub{2}]  < [\lb{1},\ub{1}]  $ & all UNSAT & $\emptyset$\\
    \textbf{(3)} & $\lb{1} \leq [\lb{2}, \ub{2}] $  & ambiguous & $\mathcal{R}[\sigma_1^{\min}]$\\
    \textbf{(4)} & $[\lb{1}, \ub{1}] \leq  \ub{2}$  & ambiguous  & $\mathcal{R}[\sigma_2^{\max}]$\\
    \textbf{(5)} & $\lb{2} \leq \lb{1}  \leq \ub{2} \leq \ub{1}  $ & ambiguous  & $\mathcal{R}[\sigma_1^{\min}] \cap \mathcal{R}[\sigma_2^{\max}]$ \\\hline
    \end{tabular}
\captionsetup{font=small}
\vspace{0.5em}
\caption{Reasoning about feasibility of family $\mathcal{F}$ using interval logic. }
\label{tab:ar_reasoning}
\vspace{-2em}
\end{table}

\paragraph{Special cases of $\varphi_c$.} Note that if $T_1=T_2$, then $\sigma_1^{\min} = \sigma_2^{\min}$: the controller minimising the reachability probability for state $s_1$ also minimises the reachability probability from state $s_2$.
Therefore, only two model checking queries are required, as opposed to four.
Finally, if $T_1 \neq T_2$ but the property quantifies over a single controller, e.g.~$\phistruc \land \exists \, \sigma: \exists \, s_1 \in \{s_1\}[\sigma], \, \exists s_2 \in \{s_2\}[\sigma]: \phiprob$, then again four controllers are computed.
% However, additional care must be taken in case \textbf{(5)}: although $\sigma_1^{\min}$ and $\sigma_2^{\max}$ may be consistent, they may be \emph{incompatible} with each other, i.e.~in some state the two controllers cannot pick the same action.
However, additional care must be taken in case~\textbf{(5)}.
If $\sigma_1^{\min}$ and $\sigma_2^{\max}$ are consistent, note that the property calls for a 1-controller, so we need $\sigma_1^{\min} = \sigma_2^{\max}$.
If this is not the case, i.e.~$\sigma_1^{\min}(s) \neq \sigma_2^{\max}(s)$ for some $s \in S$, it might happen that for one controller this action selection is not relevant to guarantee feasibility, so we can ensure $\sigma_1^{\min}(s) = \sigma_2^{\max}(s)$.
If the two controllers cannot reach an agreement for all action selections, we say that these controllers are \emph{incompatible}.

\subsection{Splitting strategy}
\label{subsec:splitting}
Consider controller $\sigma \in \Sigma^M$ provided by the model checker from which a feasible realisation $r \in \mathcal{R}$ could not have been deduced because $\sigma$ is (intra-controller) inconsistent (cases \textbf{(3)}-\textbf{(4)}, $\mathcal{R}[\sigma] = \emptyset$)\footnote{Case \textbf{(5)}, when $\sigma$ is consistent but incompatible with another controller $\sigma'$, is handled analogously, as well as when they are inter-controller inconsistent.}.
Let~$i$ be the index controller~$\sigma$ is associated with in the set of controllers to be synthesized.
$\sigma$ being inconsistent means there exists a \emph{conflicting parameter} $k \in \mathcal{K}$ associated with multiple states for which $\sigma$ selected different actions.
Formally, let $\mathcal{S}_k \coloneqq \set*{s' \in S \mid k(i,s') \simeq k}$ denote the set of states associated with $k$ and let $\mathcal{A}_k \coloneqq \set*{\sigma(s) \mid s \in \mathcal{S}_k}$ be the set of actions selected by~$\sigma$ in these states; $|\mathcal{A}_k| > 1$ iff $\sigma$ is inconsistent.

% Act(\sigma,\kp{i}{s})
% S[k(i,s)]

The goal is to partition $\mathcal{R}$ into subsets of realisations in which this conflict cannot occur, i.e.~$\sigma$ cannot be derived by the model checker.
Such a partition can be achieved by partitioning the domain $\dom{i}{s}$ of the conflicting parameter.
We construct the partition $\mathfrak{X}$ using singletons containing conflicting actions: $\mathfrak{X} = \set{ \set{a} \mid a \in \mathcal{A}_k } \cup  \set{\dom{i}{s} {\setminus} \mathcal{A}_k} $.
Let $\set{\mathcal{R}_X}_{X \in \mathfrak{X}}$ denote the corresponding partition of the set $\mathcal{R}$ of realisations where $\mathcal{R}_X \coloneqq \set*{r \in \mathcal{R} \mid r(\kp{i}{s}) \in X}$.
Finally, for arbitrary $X \in \mathfrak{X}$, consider a \emph{restriction} $M[\mathcal{R}_X]$ of MDP $M$, in which only actions from the set $X$ are available in states $\mathcal{S}_k$.
It is easy to see that $\sigma \not \in \Sigma^{M[\mathcal{R}_X]}$, indicating that over-approximation $M[\mathcal{R}_X]$ is more precise.

When $\sigma$ violates multiple structural constraints, there are multiple conflicting parameters.
To select a single parameter to be used for splitting, we employ a~heuristic inspired by~\cite{Andriushchenko022}.
The heuristic is described in Appendix~\ref{ap:metric}.

\algnewcommand\algorithmicforeach{\textbf{For each}}
\algdef{S}[FOR]{ForEach}[1]{\algorithmicforeach\ #1\ \algorithmicdo}

\setlength{\textfloatsep}{15pt}
\begin{algorithm}[t]
    \caption{Deductive syntheses using abstraction refinement (AR loop).}
    \hspace*{\algorithmicindent} \textbf{Input}: MDP $M = \mdpT$, specification $\existschedulers \in \Sigma^M \colon \phistruc \land \phistate$ where $\phistate = \existinitials \colon \phiprob$ \\
    \hspace*{\algorithmicindent} \textbf{Output}: $\osigma = (\sigma_1, \dots \sigma_n)$ that satisfies $\varphi^{struc} \wedge \varphi^{prob}$, or ``unfeasible''.
    \begin{algorithmic}[1]
    \State $\Phi \leftarrow \texttt{instantiateQuantifiers}(\phistate,\phiprob)$ \Comment{Use Def.~\ref{def:inst}}
    \State $\mathcal{R} \leftarrow \texttt{constructFamily}(M,n,\phistruc)$ \Comment{Use Def.~\ref{def:family}}
    \State $\overline{\mathcal{R}}. \texttt{push}(\mathcal{R})$ \Comment{Create a stack of (sub-)families $\overline{\mathcal{R}}$}
    \While{$\texttt{nonEmpty}(\overline{\mathcal{R}})$}
    \State $ \mathcal{R} \leftarrow \overline{\mathcal{R}}.\texttt{pop}() $
    \ForEach {$\phi \in \texttt{atoms}(\Phi)$} 
    % \Comment{Iterate over all atoms in $\Phi$}
        \State $bounds,controllers \leftarrow \texttt{analyseMDP}(M[\mathcal{R}],\phi)$ \Comment{Compute bounds wrt. $\phi$}
        \label{alg:line:model-check}
        \State  $consistency \leftarrow \texttt{checkConsistency}(controllers)$
        \State  $results(\phi) \leftarrow \texttt{intervalLogic}(bounds,consistency)$ \Comment{Apply Tab.~\ref{tab:ar_reasoning}}
        
    \EndFor
    \State $\Phi' = \texttt{replaceAndEvaluate}(\Phi, results)$
    \label{alg:line:replace}
    \If{$\Phi'\ \equiv \top$} {\Return any  $\osigma^{r}$ s.t. $r \in \mathcal{R}$} \Comment{$\mathcal{R}$ is feasible}
    \EndIf
    \If{$\Phi'\ \equiv \bot$} { \textbf{continue} } \Comment{$\mathcal{R}$ is unfeasible and thus pruned}
    \EndIf
    \If{$(\mathcal{R}' = \texttt{compose}(\Phi',results)) \neq \varnothing$}
        {\Return any  $\osigma^{r}$ s.t. $r \in \mathcal{R}'$} \Comment{SAT}
    \EndIf
    \label{alg:line:compose}
    \State $\phi \leftarrow \texttt{ambiguousAtom}(\Phi')$ 
    \State $\overline{\mathcal{R}}.\texttt{push}( \texttt{split}(\mathcal{R}, results(\phi)))$   \Comment{Use splitting wrt. $\phi$ as described in Sec. \ref{subsec:splitting}}
    \EndWhile
    \State \Return \texttt{unfeasible} 
    \end{algorithmic}
    \label{alg:AR}
\end{algorithm}
\vspace{-0.5em}
\subsection{Synthesis algorithm (AR loop)}
\label{subsec:algorithm}

Algorithm~\ref{alg:AR} summarises the proposed synthesis approach, further denoted as \emph{abstraction refinement (AR) loop}.
We first instantiate the state quantifications of $\phistate$ on $\phiprob$ (see Def.~\ref{def:inst}).
The result is $\Phi$, a Boolean combination of constraints.
Afterwards, we construct the family of realisations $\mathcal{R}$ according to Def.~\ref{def:family} and push $\mathcal{R}$ to the stack $\overline{\mathcal{R}}$ of families to be processed.
Then, the main synthesis loop starts: each iteration corresponds to the analysis of one (sub-)family of realisations.
Firstly, an iteration analyses the restricted MDP $M[\mathcal{R}]$ wrt.~each atomic property $\phi$ appearing in $\Phi$.
% namely, it enumerates the atoms $\phi$ of the formula $\Phi$ that represent the particular probabilistic constraints and check the feasibility of $\mathcal{R}$ with respect to $\phi$.
% Note that, in order to have meaningful iterations of the AR loop, available actions in $M$ are restricted according to the current $\mathcal{R}$.
This is the most computationally demanding part of the algorithm, where bounds on particular reachability probabilities (rewards) involved in $\phi$ are computed.
% In particular, two bounds are computed for standards PCTL constraints comparing a single probability (reward) with a threshold and four bounds are computed for hyper-constraints comparing two probabilities (rewards).
The produced controllers are checked for consistency, and together with bounds we use reasoning from Sec.~\ref{subsec:ar}, and in particular Tab.~\ref{tab:ar_reasoning}, to mark feasibility of $\mathcal{R}$ wrt.~$\phi$ as one of `all SAT', `all UNSAT' or `ambiguous'.
% For every atom $\phi$ the results of the analysis is stored in $results(\phi)$.
% The details of these key three steps (Line 7-9) are described in Sec.~4.1.
To decide whether $\Phi$ can be satisfied, on Line~\ref{alg:line:replace} we replace each (UN)SAT atom with $\top$ ($\bot$).
%\fra{I modified the part down here.} \milan{Okay}
If the resulting formula $\Phi'$ is trivially $\top$, then any realisation in $\mathcal{R}$ is a satisfying solution. Conversely, if $\Phi'$ is $\bot$, there is no satisfying realisation, hence $\mathcal{R}$ is discarded. 
Finally, if $\Phi'$ is neither $\top$ nor $\bot$, we try to compose a satisfying $n$-controller from individual controllers (obtained when analysing atoms of $\Phi$) by performing intersections or unions of the corresponding sets $\mathcal{R}[\sigma]$ of SAT realisations. If such composition returns an empty set, then $\mathcal{R}$ is partitioned (see Sec.~\ref{subsec:splitting}) into sub-families that are put on top of stack~$\overline{\mathcal{R}}$.

%% file: sections/5-eval.tex
\vspace{-0.5em}
\section{Experiments} \label{sec:eval}
Our experimental evaluation focuses on the following key questions:

\begin{itemize}
\item[Q1:] \emph{How does the proposed deductive approach perform in comparison to the state-of-the-art SMT based approach implemented in \textsc{HyperProb}~\cite{DobeABB21}?}
\noindent
\item[Q2:] \emph{Can the deductive approach effectively handle synthesis problems that include structural constraints?}
\noindent
\end{itemize}
 
\paragraph{Implementation and experimental setting.} We have implemented the proposed AR loop on top of the tool \textsc{Paynt}~\cite{AndriushchenkoC21} -- a tool for the synthesis of probabilistic programs.  \textsc{Paynt} takes input models expressed as extended \textsc{PRISM}~\cite{KwiatkowskaNP11} files, and relies on the probabilistic model checker \textsc{STORM}~\cite{HenselJKQV22} to analyse the underlying MDPs. All experiments were run on a Linux Ubuntu 20.04.4 LTS server with a 3.60GHz Intel Xeon CPU and 64GB of RAM. A timeout of 1 hour is enforced. All run-times are in seconds.

%\vspace{-0.5em}
\subsection*{Q1: Comparison with \textsc{HyperProb}} 
In this section, we present a detailed performance comparison of the proposed approach with the SMT encoding implemented in \textsc{HyperProb}.

We first consider the main case studies presented in \cite{DobeABB21}, namely, \textbf{TA} (timing attack), \textbf{PW} (password leakage), \textbf{TS} (thread scheduling) and \textbf{PC} (probabilistic conformance). A short description of the models and considered variants can be found in Appendix~\ref{ap:models}.
Note that the first three case studies are formulated as \emph{verification}  problems and thus the dual formulation, i.e., \emph{synthesis of a counter-example}, is considered. Also note that all considered variants are feasible, i.e. a counter-example exists as well as a valid solution for \textbf{PC}.

%\sarah{Do we describe the * more complex variants somewhere? I know we lack space, but they seem to be important as they differ from the original paper.}
%\milan{There is some info in the appendix -- maybe Franceso can add more if needed.}
%\fra{I have added a sentence in the appendix. I wouldn't say more, it would require going too much into details and explaining what was wrong with the original experiments.}

% AR -- all problems are sat <----
\small
\begin{table}[t]
\centering
\begin{tabular}{l c c c c c c c c}
\toprule
\multirow{2}{*}{ \textbf{Case Study}}  & MDP  & \multicolumn{3}{c}{AR loop} & \multicolumn{4}{c}{\textsc{HyperProb}} \\
 \cmidrule(rl){3-5}  \cmidrule(rl){6-9}
  & size & family & time & iters & vars & sub & e-time & s-time   \\
\midrule
PW $(n=4)$         & 298  & 5E21   &  ${<}1s$              & 1    &  357898  & 597  & 93   &  40   \\ % original one
{PW}$^{*}$ $(n=4)$  & 298 & 5E21    &  ${<}1s$             & 1    &  1792172 & 597  & 548  &  \texttt{TO}    \\
TA $(n=4)$          & 244 & 5E21    &  ${<}1s$             & 1    &  240340  & 489  & 63   &  491    \\ % original one
{TA}$^{*}$ $(n=4)$  & 244 & 5E21    & ${<}1s$              & 1    &  1204628 & 489  & 373  &  \texttt{TO} \\
TS ($h = (10, 20)$)   & 45 & 2E6     & ${<}1s$              & 1    &  12825   &  91  & 4     &   ${<}1s$    \\
PC ($s=0 \dots 4$)  & 20 & 3E9      & \texttt{TO}(8\%)  & 452k &  6780    &  41  & 7    &  ${<}1s$   \\
\bottomrule
\end{tabular}
\captionsetup{font=small}
\vspace{0.5em}
\caption{Comparison on \textsc{HyperProb} benchmarks. * denotes a more complicated variant not considered in \cite{DobeABB21}. For more details, see Appendix~\ref{ap:models}.} 
\label{tab:HyperProb_AR}
\vspace{-1em}
\end{table}
\normalsize

Table~\ref{tab:HyperProb_AR} reports for each benchmark the size of the underlying MDP and the main performance statistics of the proposed AR loop and \textsc{HyperProb}, respectively. For AR loop, it reports the size of the family $\mathcal{R}$, the run-time and the number of AR iterations (the number of investigated sub-families of $\mathcal{R})$. For \textsc{HyperProb}, it reports the number of variables (vars) and subformulae (sub) in the SMT encoding, together with the encoding time (e-time) and the time to solve the formula~(s-time).

We can observe that the first three case studies are trivial for the AR loop as only a single iteration is required to find a counter-example that disproves the specification. On the other hand, \textsc{HyperProb} has to construct and solve non-trivial SMT formulae and, in some cases, fails to do so within the 1-hour timeout.
The situation is very different for \textbf{PC}. \textsc{HyperProb} is able to quickly find the solution due to the manageable size and compact structure of the underlying SMT encoding. On the other hand, the AR loop fails to find a valid solution (within the timeout it explores only 8\% or the family members). A deeper investigation indicates that, in this case, the current splitting strategy is not effective.

These results are quite encouraging as they show that the proposed approach can almost trivially solve a number of relevant verification problems considered in the literature. On the other hand, they do not provide insight into the performance of our approach because the very coarse abstraction is already sufficient to find a satisfying solution. 
In the following section, we consider synthesis problems that are principally much harder for our approach.   

\vspace{-1em}
\subsubsection*{Stochastic Domination (SD)}\cite{barthe_hsu_2020} is a relation property that has been introduced for random variables. We adapt it to the reachability problem in MDPs.
We say that state $s_2$ \emph{dominates} state $s_1$ with respect to the target set $T$ if, for all controllers  $\sigma_i \in \Sigma^M$, it holds that $\reach{M^{\cv{}},s_1}{T} >  \reach{M^{\cv{}},s_2}{T}$. To verify \textbf{SD}, we consider synthesis of a counterexample, i.e. a controller proving that $s_2$ does not dominate $s_1$ formulated using the following hyperproperty: 
$$\exists \, \sigma : \forall s_1 \in \{s_a\}[\sigma] \, \forall s_2 \in \{s_b\}[\sigma] \; \,: \, \reach{M^{\ci{}},\sv{1}}{T}  \geq \reach{M^{\ci{}},\sv{2}}{T}  .$$

% SD AR
\small
\begin{table}[t]
\centering
\begin{tabular}{l c c c c c c c c c}
\toprule
\multirow{2}{*}{ \textbf{Maze}}  & \multirow{2}{*}{Feasible} & MDP  & \multicolumn{3}{c}{AR loop} & \multicolumn{4}{c}{\textsc{HyperProb}} \\
 \cmidrule(rl){4-6}  \cmidrule(rl){7-10}
  && size & family & time & iters & vars & sub & e-time & s-time  \\
\midrule
simple     &  x       & 10 & 16k   & 1        & 1.5k        & 590   & 21 & ${<}1s$ & 1          \\  
splash-1   &  $\surd$ & 16 & 1E7   & ${<}1s$       & 63(10\%) & 1424  & 33 & ${<}1s$ & 50          \\
larger-1   &  $\surd$ & 25 & 3E11  & 177      & 171k(8\%)   & 3350  & 51 & 1 & \texttt{TO}   \\
larger-2   &  x       & 25 & 7E10  & 443      & 442k          & 3350  & 51 & 1 & \texttt{TO}    \\
larger-3   &  $\surd$ & 25 & 7E10  & ${<}1s$        & 119(10\%)     & 3350  & 51 & 1 & \texttt{TO}     \\ 

splash-2   &  x       & 25 & 2E11  & 1731     & 1E6       & 3350  & 51 & ${<}1s$ & \texttt{TO}    \\ 
train      &  x       & 48 & 2E10  & 7        & 8.7k      & 11952 & 97 & 3 & \texttt{TO}    \\
% 177.27
% 443.25
% 1731.33
\bottomrule
\end{tabular}
\captionsetup{font=small}
\vspace{0.5em}
\caption{\textbf{SD} on different variants of maze problems.}
\vspace{-1em}
\label{tab:SD_AR}
\end{table}
\normalsize

We verify this specification against several variants of the problem from Example~\ref{ex:motivating_example}, where we consider a robot navigating through a maze~\cite{Norman0Z15}. Due to its faulty locomotion, after each step the robot may end up in a different location than the one suggested by the controller. We assume full observability of the maze, hence the controller uses the current state to decide the next movement direction.
Individual variants of the problem are described in Appendix~\ref{ap:models}.
Evaluation results are reported in Table~\ref{tab:SD_AR}, where we explicitly report whether the problem is feasible.
For feasible instances, the column \emph{iters} also reports the portion of the family members explored before the satisfying solution was found. Observe that \textsc{HyperProb} fails to solve a majority of the problems within the 1-hour timeout, despite the fact that the underlying SMT encoding is smaller comparing to the \textsc{HyperProb} benchmarks investigated in Table~\ref{tab:HyperProb_AR}. This is probably due to the fact that these \textbf{SD} problems are either unfeasible or have only few satisfying solutions.
On the other hand, AR loop is able to solve all \textbf{SD} variants within the timeout. Note that, although these benchmarks consider MDPs with tens of states, the individual run-times differ significantly.
This is because a specific structure of the MDP can have a huge impact on the number of iterations (splittings) required to obtain unambiguous cases.
% , although the run-times differ significantly. The run-time indeed depends on the size of the underlying family but namely on the number of iterations required to find the solutions or prune the entire family. We observe that this number heavily depends on the concrete structure of the synthesis problem and surprisingly differs significantly across the particular variants of the maze problem.  

%Although the sizes of the underlying MDPs are quite small (we consider larger problems later), and the encoding sizes are relatively small as well, we observe that \textsc{HyperProb} struggles to solve the more complicated variants, as they are either non feasible or the satisfying controllers are a tiny portion of the family. In fact, in the Case Studies of Table \ref{tab:HyperProb_AR} large portions of the families were constituted by counterexamples, hence \textsc{HyperProb} manages to synthesize one despite the larger encoding sizes. 

% SD AR explore_all
\small
\begin{wraptable}[11]{r}{0.47\textwidth}
\captionsetup{font=small}
\vspace{-1em}
\begin{tabular}{c cc  c}
\toprule
\multirow{2}{*}{ \textbf{Maze}}   & \multicolumn{2}{c}{AR loop} & feasible\\
\cmidrule(rl){2-3}
  & time & iters & instances \\
\midrule
    splash-1   & ${<}1s$        & 1233  & 1 ($\thicksim$0\%)    \\ 
    larger-1 & 4771  & 4E6 & 4($\thicksim$0\%)             \\
    larger-3   & 169      & 149k  & 2590 ($\thicksim$0\%)    \\
   
\bottomrule
\end{tabular}
\caption{Complete synthesis for the feasible variants of the \textbf{SD} problem.}
\label{tab:SD_AR_explore_all}
\end{wraptable}
\normalsize

To further emphasise the capabilities as well as the limitations of the proposed approach, we consider the \emph{complete synthesis} problem: instead of synthesizing a single satisfying controller, we look for \textbf{all} controllers that meet the specification, which requires exploration of the entire family. Note that this problem is closely related to a construction of \emph{permissive controllers} or \emph{safety shields} in the context of safe reinforcement learning~\cite{10.1007/978-3-642-54862-8_44,jansen_et_al:LIPIcs:2020:12815}.
%The complete synthesis  requires exploration of the entire family and typically the abstraction-based pruning is less efficient comparing to the unfeasible instances. 
Table~\ref{tab:SD_AR_explore_all} shows the performance of the AR loop as well as the number of feasible instances. As expected, the complete synthesis for feasible instances is significantly harder.
Moreover, we observe that abstraction-based pruning is typically less efficient comparing to the unfeasible instances. For the variant larger-1, only 86\% of the family members were explored in the 1-hour timeout (the entire exploration took 80 minutes). Complete synthesis is also very challenging for the SMT approach since it requires multiple calls to the SMT solver, where each subsequent call rules out the previously found solution. \textsc{HyperProb} currently does not support complete synthesis.

%\emph{Conclusion:} Our experimental evaluation clearly demonstrates that the proposed approach significantly outperforms \textsc{HyperProb} and is able to scale to more complicated synthesis problems.

%the state-of-the-art SMT-based approach inplemented in \textsc{HyperProb}.

\subsection*{Q2 Synthesis under structural constraints}
To investigate the performance of the proposed approach on more complicated synthesis problems that include structural constraints on the resulting controllers, we extend the robotic scenario from Example~\ref{ex:motivating_example}. We consider \emph{probabilistic noninterference}~\cite{GoguenM82a} and \emph{opacity} ~\cite{0044NP20}, both ensuring location privacy with respect to attacker observability. In both cases we formulate the problem within a planning scenario in a maze. To demonstrate the scalability with respect to the complexity of the underlying MDP, we consider two larger variants of the maze, (see the MDP sizes in Table~\ref{tab:Q2}) and a more complicated objective.  The maze is equipped with checkpoint locations and the robot earns a reward when repeatedly passing through at least two of these locations. Hence, a good strategy makes the robot move back and forth among two or more checkpoints. There are no traps, but every move has a small chance of unrecoverable error. Additionally, we assume \emph{partial observability} - the robot can only observe the available directions (up, down, left, right). This is encoded using intra-controller constraints $\obs{S'}{\sigma}$. 

% ProbNI
\subsubsection{Probabilistic noninterference} is formulated as a 2-controller synthesis problem with two different initial states that need to be kept private. We consider the specification containing two PCTL properties requiring that both controllers achieve a reward exceeding a minimum threshold $\lambda$ and a hyperproperty requiring that the achieved values are almost the same (the attacker observes the final rewards). Additionally, we define a set of sensitive states where both controllers have to agree on their actions, encoded using inter-controller constraints $\same{s}{\Sigma}$.
For both considered variants of the maze (larger-4 and larger-5), there is no pair of memoryless controllers that would satisfy the specification with a non-zero threshold.
In Table~\ref{tab:Q2} (left) we report experiments where threshold $\lambda$ is set to 0, already making the problem unfeasible.

Furthermore, we enhance both MDPs with a 1-bit memory (denoted by suffix~+mem).
A controller can now properly use the memory to distinguish between otherwise indistinguishable states.
With this extension, it is possible to find 2-controllers satisfying probabilistic noninterference with non-zero rewards. 
In the corresponding experiments in Table~\ref{tab:Q2}, the reward thresholds $\lambda$ are set close (but not equal) to the respective maximal values.
This makes the PCTL part of the specification satisfiable, but the overall specification unfeasible. This allows us to inspect how the hyperproperty is leveraged to explore the entire search space. The results show that adding the 1-bit memory enormously enlarges the family size; however, the AR loop is still able to fully explore the family in a reasonable time. Exploring the family of 2-bit controllers is not tractable within the 1-hour timeout. 

%ProbNI and Opacity with AR; for PRobNI, all experiments are UNSAT.
% larger-4 is Small Grid; larger-5 is Medium Grid
\small
\begin{table}[t]
\centering
\begin{tabular}{c c  ccc  cccc}
\toprule
 \multirow{2}{*}{\textbf{Maze}} & MDP & \multicolumn{3}{c}{Noninterference} &  \multicolumn{4}{c}{Opacity} \\
\cmidrule(rl){3-5} \cmidrule(rl){6-9}
 & size & family  & time & iters &  family  & time & iters & distance \\
\midrule

larger-4 & 434   & 221k   & 1        & 337  & 3E9  & ${<}1s$    & 96   & 6/13 \\ 
larger-5 & 1260  & 3E6    & ${<}1s$        & 7    & 5E11 & ${<}1s$    & 43   & 8/15 \\ 
larger-4+mem  & 866   & 8E20   & 1551      & 110K  & 2E38 & 193  & 13k  & 32/32 \\  
larger-5+mem  & 2520  & 1E25   & 481      & 16k  & 2E46 & 309  & 9k   & 22/38 \\
                      
\bottomrule
\end{tabular}
\vspace{0.5em}
\captionsetup{font=small}
\caption{\textbf{Probabilistic Noninterference} and \textbf{Opacity} specifications for more complex variants of the maze problem.}
\vspace{-1em}
\label{tab:Q2}
\end{table}
\normalsize

\subsubsection{Opacity} is a hyperproperty, used in non-probabilistic robot planning, specifying that for a given state there exist two different paths that both give identical observations to the attacker and reach the planning goal.
We adapt opacity to our setting in the following way.
For a given initial state, we seek for two controllers that i) achieve an expected reward above a threshold (performance requirement), ii) achieve almost the same reward visible to the attacker (anonymity requirement) and iii) take as many mutually different choices as possible.
If the two controllers are indistinguishable for the attacker and yet do not agree on any choice, then anonymisation is fully guaranteed. We encode the last objective as an optimisation over the intra-controller characteristic, further denoted as \emph{distance}, where we maximise the number of different choices.

In order to effectively search for the \emph{optimal} $n$-controller wrt.~such a specification, we need to completely explore the family $\mathcal{R}$. We extend the AR loop to use the distance of the current optimum to improve the pruning strategy: if the current sub-family does not include improving solutions wrt.~currently optimal distance, it is promptly pruned without analysing the MDPs wrt.~performance and anonymity properties. This way, we extend the deductive reasoning over the families using an additional structural constraint that iteratively changes during the synthesis process.

%This can be formulated with a slight variation of the presented intra-controller constraint $\same{s}{\Sigma}$. Instead of forcing the two controllers to take the same action in some states, we allow them to make choices freely, but the AR loop is adapted to keep track of such constraint in the following iterative way. Recall that an optimization objective always requires to explore the entire family. 

%When the AR loop finds a satisfying solution for the remaining specification, it records the difference between the two controllers of the solutions. This difference is then used as a threshold: when analysing subsequent families, it evaluates, besides the quantitative constraints, the maximum amount of different choices such family can take, and if lower than current threshold, it prunes it. When a new satisfying solution with more difference is found, the threshold is updated accordingly.

%As before, we require that both controllers achieve (almost) the same expected reward that is above a certain threshold. 
%We again use reward thresholds requiring a nontrivial exploration with respect to the hyperproperty and the maximisation objective. If the two controllers do not agree on any choice, then \emph{anonymization} is fully guaranteed. 

Table~\ref{tab:Q2} (right) reports the results for memoryless and 1-bit controllers, respectively. The last column reports the obtained maximal distance and the number of choices that need to be synthesised per controller.
Note that, due to partial observability, there are fewer choices than states. We see again that adding memory makes the synthesis problem much harder. On the other hand, the memory not only allows us to significantly increase the threshold on the reward, but also improves the achieved distance, i.e. the degree of anonymisation. For the variant larger-4+mem, we synthesized a pair of controllers that differ in all 32 choices, thus ensuring full anonymisation.

%% file: sections/conclusions.tex
\section{Conclusions}
\label{sec:conclusion}
We proposed a deductive controller synthesis procedure for MDPs with respect to probabilistic hyperproperties 
expressed in HyperPCTL enriched with additional structural constraints on the resulting controllers. Our approach builds on a symbolic representation of the family of the controllers and an abstraction refinement strategy allowing for an effective exploration of the families. A detailed experimental evaluation demonstrates that our approach considerably outperforms \textsc{HyperProb}~\cite{DobeABB21}, a state-of-the-art tool based on SMT reasoning. In particular, our approach scales to more complicated synthesis problems where \textsc{HyperProb} times out, and, for the first time, effectively supports in the synthesis procedure structural constraints on the resulting controllers. As future work, we plan to improve our approach by investigating counterexamples for probabilistic hyperproperties. 

% Way too long
% Probabilistic hyperproperties is a class of system level properties that specify a relation between independent probabilistic computation trees. This class of requirements  is suitable to express probabilistic quantitative information-flow properties such as probabilistic noninterference and information-leakage.  We propose a new deductive controller synthesis procedure for MDPs satisfying probabilistic hyperproperties. 
% As specification language we consider HyperPCTL that we enrich for the first time with  with structural constraints that allows to handle partially observable MDP and the possibility to synthesize strategies that agree on a common action for a given state.  We compare our approach with the SMT-based approach implemented in the model checking tool \textsc{HyperProb}.  Our experimental results are very encouraging outperforming \textsc{HyperProb} on several benchmarks. As future work we plan to improve our synthesis approach by investigating the notion of counterexample for probabilistic hyperproperties and by using the counterexample to guide the synthesis also inductively.